# Even quantum pigeons may thrive together

## A note on "the quantum pigeonhole principle"

B. E. Y. Svensson

*Theoretical High Energy Physics,*
*Department of Astronomy and Theoretical Physics,*
*Lund University, Sölvegatan 14, SE-22362 Lund, Sweden*

Abstract

The findings in the paper "The quantum pigeonhole principle and the nature of quantum correlations", (arXiv 1407.3194), by Aharonov, Colombo, Popescu, Sabadini, Struppa and Tollaksen are scrutinized. I argue that some of the conclusions in the paper are ambiguous in the sense that they depend on the precise way one defines correlations, and that the "first experiments" the authors suggest has little if any bearing on their main theses. The far-reaching conclusions the authors reach seems, therefore, premature.

## I. INTRODUCTION

The paper "The quantum pigeonhole principle and the nature of quantum correlations", (arXiv 1407.3194), by Aharonov, Colombo, Popescu, Sabadini, Struppa and Tollaksen [1] contains some surprising conclusions. Even journals like Physics World [2] and New Scientist [3] have paid attention to the paper, hailing its result as "… could have immense implications " [2] and "This is a surprising breakthrough" [3].

A direct quote from the paper reads

> "The pigeonhole principle: 'If you put three pigeons in two pigeonholes at least two of the pigeons end up in the same hole' …. Here, however, we show that in quantum mechanics this is not true! We find instances when three particles are put in in two boxes, yet no two particles are in the same box. " [1]

The paper also contains some general remarks on correlations and asserts that the results "… shed new lights on the very notion of separability and correlations in quantum mechanics and on the nature of interactions. It also presents a new role for entanglement ….".

My present note provides some further aspects on the results in [1]. In particular, I show that, in quantum mechanics (QM), breaking the pigeonhole principle in the case of a three-particle system depends crucially on how you interrogate the system; you can easily find examples where a QM pigeonhole principle is valid. I also comment on the experiment the authors suggest to illustrate the purported effects.

The physical picture employed in [1] uses Mach-Zehnder interferometers. In an appendix, I point out that the formalism in [1] may as well describe spin-½ particles. Another appendix contains some comments to the remarks in [1] on the different type of correlations that occur in "global" *versus* "detailed" measurements (I will explain the meaning of these terms in Appendix B below): I point out that these remarks are not specific to three- or multiparticle states. Indeed, the difference is precisely the important difference in QM between coherent and incoherent processes.

As the authors of [1] do – more or less explicitly – I will approach the problem by using two methods. The first is what I call the ABL approach, *i.e.,* employing the results in the paper by Aharonov, Bergmann and Lebowitz [4]. They deduced a general expression for the probability, which I call the "ABL probability", of finding a system[1] in a given intermediate eigenstate, say $|a>$ with eigenvalue $a$, conditioned on the system being prepared in a given initial state and ending up in a given final state. The second method is based on weak values, as used *e.g.* by Vaidman [5][2] , to formulate a weak measurement criterion for the presence of an intermediate state $|a>$ in terms of the weak value of the projection operator $\Pi_a = |a><a|$. – As a kind of short-hand writing, I shall allow slightly ambiguous or even dubious phrases like "the system can be found in the intermediate state | a >", "the system has the (intermediate) value $a$" and the like to mean that the ABL probability (respectively the corresponding weak value) for finding the value $a$ is different from zero.

---

[1] As a kind of short-hand writing, I shall in the sequel allow slightly inappropriate or even dubious phrases like "the system can be found in the intermediate state | a >", "the system has the (intermediate) value $a$" and the like to mean that the ABL probability (respectively the corresponding weak value) for finding the value $a$ is different from zero.

[2] This paper may also be consulted for further references on the weak measurement-weak value approach. Another presentation of this approach is [6]

## II. THE ABL APPROACH

The authors of [1] study three-particle states in which each particle has two different states, $|L>$ and $|R>$, symbolizing the left and right arm of a Mach-Zehnder interferometer (MZI). (In the paper, they are described as each particle – alternatively dubbed "pigeons" – being in the $L$ respectively the $R$ "box".) The three-particle system is prepared – "preselected" – in the direct-product state

$$|in> = |+>_1 |+>_2 |+>_3 , \qquad (1)$$

where

$$|+>_1 = (|L>_1 + |R>_1)/\sqrt{2} , \qquad (2)$$

and similarly for particles *2* and *3*.

This state is now subjected to an "ABL analysis" [4]. That is, one is interested in what can be said about the system at an intermediate time in its evolution from the preselected state to a final, "postselected", state $|f>$, assuming there is nothing but free time-evolution between the different preparations/measurements. The authors consider this postselected state to be another direct-product state, *viz.*,

$$|f> = |+i>_1 |+i>_2 |+i>_3 , \qquad (3)$$

where

$$|+i>_1 = (|L>_1 + i|R>_1)/\sqrt{2} , \qquad (4)$$

with a similar notation for particles *2* and *3*.

The aim of the paper [1] is to investigate correlations between different states of the particles at an intermediate time, *i.e.*, properties of the three-particle system at a time in its evolution after the preselection but before the postselection. A particular question is whether any two of the three particles can be in identical MZI arms (or "boxes").

It is therefore convenient to introduce the projection operator

$$\Pi_1^L = |L>_1 {}_1<L| , \qquad (5)$$

with a similar definition for the other combinations of "boxes" ( $L, R$ ) and "particles" ( $1, 2, 3$ ). They are combined into correlation projectors like

$$\Pi_{12}^{same} = \Pi_1^L \Pi_2^L + \Pi_1^R \Pi_2^R, \tag{6}$$

and

$$\Pi_{12}^{diff} = \Pi_1^L \Pi_2^R + \Pi_1^L \Pi_2^R. \tag{7}$$

There are four more analogous projection operators corresponding to the two other pairings of the particle numbers *1*, *2* and *3*. Such projectors can be used to find out whether the particles *1* and *2* are in identical boxes (for $\Pi_{12}^{same}$), respectively different boxes (for $\Pi_{12}^{diff}$). I shall also have use for a few other such multiparticle correlation projectors, in particular the three-particle projector

$$\Pi_{123}^{same} = \Pi_1^L \Pi_2^L \Pi_3^L + \Pi_1^R \Pi_2^R \Pi_3^R, \tag{8}$$

when I am interested in whether all three particles occupy identical boxes.

The intermediate properties of the system may be investigated by using "ABL probabilities" [4]. For example, the probability of finding particles *1* and *2* in identical boxes, given $|in>$, eq. (1), as the preselected state and $|f>$, eq. (3), as the postselected one, is proportional to the absolute square of the "ABL amplitude" $<f | \Pi_{12}^{same} | in>$. − Here, I invoke the natural connection between a projection operator and a question (or a statement) that can be given a yes or no answer (respectively assigned a value of true or false). For example, the question "Can the particles *1* and *2* be found in identical boxes?" is represented by the projector $\Pi_{12}^{same}$.

Since I shall only be interested in the vanishing or non-vanishing of this and other ABL probabilities, I shall only exhibit whether the corresponding *amplitudes* vanish of not. To avoid too much writing, I shall also say that a particle pair is (or the particles themselves are) *positively (*respectively *negatively) correlated* if the particles are found in identical (respectively different) boxes.

In fact, the amplitude $<f | \Pi_{12}^{same} | in>$ does vanish, so the pair (*1*, *2*) is negatively correlated: the probability of finding this pair in identical boxes at an intermediate time is zero. From the fact that the two projectors $\Pi_{12}{}^{same}$ and $\Pi_{12}{}^{diff}$ are orthogonal and sum to unity,

$$\Pi_{12}{}^{same} + \Pi_{12}{}^{diff} = \mathbb{1}_{12} , \tag{9}$$

it follows that the two particles can be found in different boxes only.

Due to the symmetry of the pre-and postselected states, the same conclusion applies to the particle pairs (*2*,*3*) and (*3*,*1*). This is what is called "the quantum pigeonhole principle" in [1].

But what about a joint statement like "The two pairs, (*1*, *2*) and (*2*,*3*), are both positively correlated."? As no pair is positively correlated in itself, it seems as the statement is false. But QM logic shows some surprises. Namely, before one can answer such a question, the QM rules require us to specify the corresponding projector. It is given by $\Pi_{12}{}^{same} \Pi_{23}{}^{same} = \Pi_{123}{}^{same}$. And, lo and behold, this operator has non-vanishing ABL amplitude! So while it is true that none of the pairs are positively correlated internally, there is a non-vanishing probability for jointly finding both pairs positively correlated. Moreover, the question "Are the particles in either the (*1*, *2*) or the (*2*,*3*) pair positively correlated?" cannot even be given a yes/no answer: it would be represented by the operator $\Pi_{12}{}^{same} + \Pi_{23}{}^{same}$ which, however, is not a proper projector (it is not idempotent, *i.e.*, it is different from its square) (*c.f.* [7]).

There are more surprises to come. To see one example, let me first note that an operator like $\Pi_{12}{}^{same}$ answers the question whether the pair (*1*, *2*) is positively correlated *irrespective of the third particle.* One could, for example, instead ask if that pair is positively correlated when particle *3* is negatively correlated with respect to either of *1* or *2*. Then, one should employ a projector like

$$\Pi_{12,3}^{s,d} = \Pi_1^L \Pi_2^L \Pi_3^R + \Pi_1^R \Pi_2^R \Pi_3^L . \tag{10}$$

And this has a non-vanishing ABL amplitude. The same applies if one asks for the third particle to be positively correlated with the particles in the pair, since that is nothing but $\Pi_{123}{}^{same}$, eq. (8).

One might even argue that the choice (10) is more motivated than the choice (6); at least, it has in a sense "better" properties than (6) . What I mean is that the projector $\Pi_{12,3}^{s,d}$, together with the two other analogous projectors $\Pi_{23,1}^{s,d}$ and $\Pi_{31,2}^{s,d}$, are orthogonal (projects onto orthogonal subspaces of the three-particle Hilbert space), a property not shared by, *e.g.*, $\Pi_{12}^{same}$ and $\Pi_{23}^{same}$. Together with $\Pi_{123}^{same}$, they even form a resolution of the identity for the three-particle Hilbert space, fulfilling the condition of "defining a measurement" [7].

Summing up, and expressed in pigeonized language, the situation for the QM pigeons is such that if two of them do not care where the third pigeon is, they stay in separate boxes. But all three pigeons have nothing against squeezing into the same box. Moreover, one pair may very well thrive together provided a second pair also does. And if two pigeons do abhor being in the same box as the third one, they could very well stick together, too.

Whether this situation is paradoxical or not is up to anyone to decide. But it very clearly shows that it is important in QM to specify how to interrogate the system under study, what questions to ask, how a question is formulated, and whether a particular question can legitimately be posed (*c.f.* [7]).

### III. AN APPROACH USING WEAK VALUES

Could one invoke weak measurement and weak values to elucidate the issues involved here?

Vaidman [5] in particular has emphasized that weak values of projection operators can be used to trace whether the system under study could have been in the state corresponding to that projection operator. In other words, the weak value of a projection operator $\Pi_a = |a\rangle\langle a|$ could be used instead of ABL probabilitites to answer questions like "Could the system at an intermediate time have been in the state $|a\rangle$?".

One advantage of weak values is that weak measurements obey a slightly different, more classical-like logic than what ordinary, projective measurements do. In particular, as has been pointed out by several proponents of the weak value idea (see, *e.g.*, [6] and references therein), nothing forbids essentially simultaneous weak measurements even of non-commuting operators to be performed. And questions that cannot legitimate be posed for projective measurements may have an answer in a weak measurement approach. To take one concrete example: while, as pointed out above, an operator like $\Pi_{12}^{same} + \Pi_{23}^{same}$ does not correspond to a legitimate question for projective measurements, the adding up of the corresponding weak values, $(\Pi_{12}^{same})_{weak} + (\Pi_{23}^{same})_{weak}$, is not illegitimate.[3]

It facilitates the argument to note that, for projection operators, the ABL amplitude is the numerator of the corresponding weak value. So all statements in the previous section regarding the vanishing or not of an ABL amplitude can be directly translated to the vanishing or not of the corresponding weak value.

However, this does not solve the main puzzle. One is still caught in the QM logic of, *e.g.*, having only negatively correlated particle pairs when no regard is paid to the third particle, while two particles could be positively correlated whenever none of them is positively correlated with the third particle, or if all three particles are positively correlated. From a classical viewpoint such state of affairs might seem contradictory. In QM they depend, *i.a.*, on how exactly one formulates the conditions under which the measurements are made.

## IV. "A FIRST EXPERIMENT"

The authors of [1] suggest a "first experiment" to test their ideas. In the first paragraph of the section in their paper which treats this issue they write:

> "Consider again three particles and two boxes. Let the particles interact with each other by bipartite short-range interactions, i.e., any two particles interact when they are in the same box and do not interact otherwise. Then, as there are three particles and only two boxes *we expect* that always at least two of the particles should interact. But due to our pigeonhole effect, this is not so, as shown in the following experiment." (my italics).

That expectation could be defended based on classical physics but it is not supported in QM.

In fact, the authors are now slightly switching gears. The issue is not any longer to find out whether a particular intermediate state is present or not. Instead, they now focus on a transition type experiment in which one prepares a system in the preselected state, let it interact via an interaction Hamiltonian and then single out for acceptance merely those final state that are triggered by a detector sensitive only to the postselected state. A typical example of such transition experiments is a scattering experiment.

---

[3] As I have argued in [8], one should be aware of its interpretation, though, for example how to interpret its detailed numerical value, However, if one disregards its precise numerical value, one may interpret the vanishing of not of the weak value of an appropriate projection operator onto a state as the intermediate presence or not of the system in that state (provided it is applied properly [9]).

It is well known that in this type of experiment all kinds of interference and selection effects could be at play. As a consequence, the vanishing or not of a *transition amplitude* does not imply anything about the presence of not of a certain intermediate state.

As a concrete example, the authors propose an interaction Hamiltonian involving pairwise interaction, but only between particles in a pair that are positively correlated: the interaction Hamiltonian is a linear combination of the pair-correlation operators $\Pi_{12}^{same}$, $\Pi_{23}^{same}$ and $\Pi_{31}^{same}$. With $|in>$, eq. (1), as initial and $|f>$, eq. (3), as final state, the transition matrix element of the Hamiltonian then vanishes. But this cannot in any way be interpreted as particles (or states) being "present" or not at an intermediate time; rather, it is an example of what I just mentioned, *viz.*, it may easily be explained as interference or as a selection effect.

In fact, one may even think of combining a transition type experiment with a weak measurement probe. To take one example, let the initial state in the three-particle case under study here be subjected to a weak measurement of, say, the projection operator $\Pi_{12}^{same}$ *before* the system interacts. It follows from the discussion above that one would then find its weak value to be zero. In the weak value logic, this implies that no such pair will ever reach any interaction!

And even if one were to subscribe to the authors' arguments, the situation is not clear-cut due to the ambiguity in the choice of correlations: why choose a Hamiltonian built from $\Pi_{12}^{same}$, eq. (6), *etc.*, instead of, *e.g.*, $\Pi_{12,3}^{s,d}$, eq. (10), *etc.*? In fact, a two-body interaction proportional to $\Pi_{12,3}^{s,d}$, *etc*, would give a non-vanishing transition amplitude. (One might object that including conditions on the third particle is not a pure two-particle interaction. But one does so also in the Hamiltonian proposed in [1]: since in fact $\Pi_{12}^{same} \equiv \Pi_{12}^{same} \times \mathbb{1}_3$, there is an operator factor of unity for the third particle implicit in this two-particle correlation.)

In sum, the arguments based on transition/scattering experiments are not relevant to the findings of the previous sections of the paper.

## IV. SUMMARY AND CONCLUSIONS

The conclusions reached in [1] are vulnerable to some criticism.

Firstly, the result regarding pairwise correlations are not unambiguous. If one asks for the presence of correlations within a pair of particles in the three-particle state considered, it is important to tell which role the third particle is supposed to play. If one does not specify its state, then the conclusion drawn by the authors of negative correlation (the particles in the pair are in different states or "boxes") follows. However, the pair could be positively correlated (being in identical one-particle states, irrespective of which) if one specify the third particle to be in a definite state, either identical to that for the particles in the pair or in the different one. These conclusions are reached within an ABL approach as well as within a weak measurement-weak value approach.

Secondly, the authors of [1] suggest an experimental realization to elucidate their theoretical findings. They consider a transition/scattering experiment and assume an interaction Hamiltonian with only two-particle interactions build from their pairwise correlation projectors. This, however, switches the focus away from the "presence" of certain correlations (as measured in either of the ABL or weak value approaches): a transition experiment, typified by a scattering experiment, has little if any relevance to the "presence" type of experiment. Besides, one may raise the same, ambiguity-based objection to the choice of interaction Hamiltonian as I did in the previous paragraph for the pairing operators.

## ACKNOWLEDGEMENT.

I thank the participants in the Okazaki October 2014 workshop on weak values, in particular Ruth. E. Kastner and Yutaka Shikano, for stimulating discussions in an initial phase of the work reported here.

## APPENDIX A. SPIN REALIZATION OF THE PIGEON FORMALISM

There are other physical realizations of the formalism developed in [1] than the case with particles – "pigeons" – each in a separate MZI. In fact, the formalism is particularly well tailored for spin-½ particles.

To see this, let $|\Uparrow>$ ( $|\Downarrow>$ ) denote the z-direction spin-up (spin-down) states of a spin-½ particle and identify them with $|L>$ (respectively $|R>$) in the notation of section II above. Then, in a hopefully self-explanatory notation, the states

$$|x, \pm> := (|\Uparrow> \pm |\Downarrow>)/\sqrt{2} \quad \leftrightarrow \quad |\pm> = (|L> \pm |R>)/\sqrt{2}, \tag{A1}$$

$$|y, \pm> := (|\Uparrow> \pm i |\Downarrow>)/\sqrt{2} \quad \leftrightarrow \quad |\pm i> = (|L> \pm i |R>)/\sqrt{2}, \tag{A2}$$

are, respectively, the eigenstate of the spin projection in the x- respectively the y-direction. Furthermore, the correlation operators become, *e.g.*,

$$\Pi_1^{\Uparrow} = |\Uparrow>_{1\ 1}<\Uparrow| \quad \leftrightarrow \quad \Pi_1^{L} = |L>_{1\ 1}<L|, \tag{A3}$$

and

$$\Pi_{12}^{same} = \Pi_1^{\Uparrow} \Pi_2^{\Uparrow} + \Pi_1^{\Downarrow} \Pi_2^{\Downarrow}. \tag{A4}$$

## APPENDIX B. "THE NATURE OF QUANTUM CORRELATIONS"

One section of the paper [1] is devoted to some aspects of quantum correlations in multiparticle states more generally. The essential finding in that section is that "there is a significant difference between correlations that can be observed when we measure particles separately and when we measure them jointly".

Indeed. But it might be worth pointing out that this is a more general feature than one linked to multiparticle systems. In fact, what the authors show is that there is a difference in the ABL probabilities when one makes a "separate " or "detailed" measurement" as it is called in [1], involving an incoherent sum, *versus* a "joint" or "global" measurement, involving a coherent sum.

In more detail, let me consider a general QM system and assume one is interested in the ABL probability of an intermediate result $a_i$ for an observable $A$. Suppose further that one's interest is not limited to one particular such eigenvalue but rather to any eigenvalue in a set $\alpha = \{a_i\}$ of eigenvalues of $A$, with the index $i$ ranging from 1 to some number $N > 1$.

One may think of two different approaches to such task:

"Detailed measurement". Measure the intermediate probability of getting the value $a_i$ for each $i$ separately and sum these probabilities for all eigenvalues $a_i$ in $\alpha$. This is illustrated by the treatment in eqs. (14-15) in [1].

"Global measurement". Arrange the experiment to measure directly the intermediate probability of finding any of the eigenvalues $a_i$ belonging to $\alpha$. This corresponds to measuring the projection operator

$$\Pi_\alpha = \Sigma_{i=1}^{N}\ \Pi_i \qquad (B1)$$

with $\Pi_i = |a_i\rangle\langle a_i|$, as exemplified by the expressions in eq. (16) in [1].

To be even more specific, let me consider the crucial entity, *viz.*, the probability *prob ( f, $a_i$ / in )* of finding an intermediate value $a_i$ *and* a final state $|f\rangle$, given $|in\rangle$ as the initial state. (It is this entity that, via Bayes theorem, gives the ABL probability [4].). Then, for a "detailed" measurement, one is interested in

$$\text{"detailed probability"} = \Sigma_{i=1}^{N}\ prob(f, a_i/in) =$$
$$= \Sigma_{i=1}^{N}\ |\langle f|s_i\rangle|^2 |\langle s_i|in\rangle|^2 = \Sigma_{i=1}^{N}\ |\langle f|\Pi_i|in\rangle|^2, \qquad (B2)$$

which involves an incoherent sum. For a "global" measurement, on the other hand, the interesting probability is

$$\text{"global probability"} = prob(f, \Pi_\alpha \text{ has eigenvalue} = +1/in) =$$
$$= |\langle f|\Pi_\alpha|in\rangle|^2 = |\Sigma_{i=1}^{N}\ \langle f|\Pi_i|in\rangle|^2, \qquad (B3)$$

which indeed involves (the square of) a coherent sum.

It is no surprise that you will get different results in these two ways of approach. Whether this is a revolutionary discovery or not is up to anyone to decide. Basically, anyhow, the difference is nothing but the fundamental difference in QM between incoherence and coherence.

In the same section of [1], the final paragraph states "Finally, and most importantly, we note that the global measurement is a measurement of an operator with entangled eigenstate … ". I just note that the eigenstates to the relevant correlation operator $\Pi_{12}^{same}$ are degenerate, so there is not just one (unique) set of eigenstates. In fact, the unentangled states $|L>_1|L>_2$ and $SP|R>_1|R>_2$ are as good eigenstates of $\Pi_{12}^{same}$ as is an entangled state like $(|L>_1|L>_2 + |R>_1|R>_2)/\sqrt{2}$.